\documentclass[]{aastex631}
\usepackage{amsmath}
\usepackage{graphicx}
\usepackage{threeparttable}
\usepackage{txfonts}
\usepackage{soul}

\shorttitle{FRBs as EM Counterpart to EMRIs}
\shortauthors{Li et al.}

\begin{document}

\title{Fast Radio Bursts: Electromagnetic Counterparts to Extreme Mass Ratio Inspirals}

%%
%% While authors can be grouped inside the same \author and \affiliation
%% commands it is better to have a single author for each. This allows for
%% one to exploit all the new benefits and should make book-keeping easier.
%%
%% If done correctly the peer review system will be able to
%% automatically put the author and affiliation information from the manuscript
%% and save the corresponding author the trouble of entering it by hand.

%\correspondingauthor{August Muench}
%\email{greg.schwarz@aas.org, gus.muench@aas.org}

\correspondingauthor{Fa-Yin Wang}
\email{fayinwang@nju.edu.cn}
\author[0009-0007-3326-7827]{Rui-Nan Li}
\affiliation{School of Astronomy and Space Science, Nanjing University Nanjing 210023, People's Republic of China}
\author[0000-0002-2171-9861]{Zhen-Yin Zhao}
\affiliation{School of Astronomy and Space Science, Nanjing University Nanjing 210023, People's Republic of China}

\author{Zhifu Gao}
\affiliation{Xinjiang Astronomical Observatory, Chinese Academy of Sciences, Urumqi, 830011, Xinjiang, People's Republic of China}

\author[0000-0003-4157-7714]{Fa-Yin Wang}
\affiliation{School of Astronomy and Space Science, Nanjing University Nanjing 210023, People's Republic of China}
\affiliation{Key Laboratory of Modern Astronomy and Astrophysics (Nanjing University) Ministry of Education, People's Republic of
	China}
\affiliation{Purple Mountain Observatory, Chinese Academy of Sciences, Nanjing 210023, China}

\begin{abstract}

Recent observations discovered that some repeating fast radio bursts (FRBs) show a large value and complex variations of Faraday rotation measures (RMs). The binary systems containing a supermassive black hole (SMBH) and a neutron star (NS) can be used to explain such RM variations. Meanwhile, such systems produce low-frequency gravitational wave (GW) signals, which are one of the primary interests of three proposed space-based GW detectors: the Laser Interferometer Space Antenna (LISA), Tianqin and Taiji. These signals are known as extreme mass ratio inspirals (EMRIs). Therefore, FRBs can serve as candidates of electromagnetic (EM) counterparts for EMRI signals. In this letter, 
we study the EMRI signals in this binary system, which can be detected up to $z\sim0.04$ by LISA and Tianqin for the most optimistic case. Assuming the cosmic comb model for FRB production, the total event rate can be as high as $\sim1$ Gpc$^{-3}$ yr$^{-1}$. EMRI signals associated with FRBs can be used to reveal the progenitor of FRBs. It is also a new type of standard siren, which can be used as an independent cosmological probe. 
\end{abstract}


\section{Introduction} \label{sec:intro}
Fast Radio Bursts (FRBs) are bright pulses with large dispersion measures (DMs) at radio frequency, with millisecond-duration randomly occurring at cosmological distances \citep{6,13,Zhang2022}. FRB was first discovered in 2007 \citep{76}, and some FRBs were observed to be active repeatedly \citep{7}. Whether there are two intrinsically different categories, repeating one and non-repeating one, has not been confirmed yet. Up to now, over 600 FRBs have been reported \citep{8}. However, their physical origins are still poorly known. It has been speculated that the central engines of FRBs are magnetars \citep{Popov2013,Kulkarni2014,Murase2016,Katz2016,Metzger2017,Wang2017,Beloborodov2017,Lu2018,Yang2018,Wadiasingh2019,Wang2020}. The discovery of FRB 20200428 produced by Galactic magnetar SGR J1935+2154 confirmed that at least some FRBs are from magnetars \citep{54,8}. %Among the population of neutron stars, about ten percent of neutron stars are expected to be magnetars \citep{56}, and population synthesis and observations both suggest a fraction of magnetars can stay in bound binaries \citep{55}. 

The variations of the RM and DM can provide us with important information about the surrounding environment of FRBs. Long-term monitoring of repeating FRBs revealed that the majority of them show large RM variations, suggesting them in dynamical magneto-ionic environments \citep{Mckinven2023}. FRB 20121102 has a large RM ($\sim10^{5}~\rm{rad}~m^{-2}$) \citep{75}, which decreased by 30\% in about two years \citep{77} and 70\% till to 2023 \citep{Feng2023}. It may be caused by the wind nebula of a young magnetar \citep{86,Zhao2021b}, the ejecta from the progenitors (massive stars or compact binary mergers \citealt{Piro2018,87}), or the outflow from an SMBH \citep{90,88,28}. FRB 20201124A showed a short-time variation of RM \citep{Xu2022} and the first local RM reversal \citep{Wang2022}. A magnetar/Be star binary is promising to explain the unusual features of FRB 20201124A \citep{Wang2022,Lu2023}, including the RM variation and reversal, depolarization and Faraday conversion.
FRB 20190520B was found to reside in a dwarf galaxy at redshift $z=0.241$. It has a lager host DM $\sim 900~\rm{pc}~\rm{cm}^{-3}$ \citep{71}, which may be related to a young supernova remnant \citep{Zhao2021b,93}. FRB 20190520B showed a large value of RM and the RM reversal \citep{Anna-Thomas2023}, which can be explained in massive binary models \citep{Wang2022,Zhao2023,Anna-Thomas2023}. FRB 20180301A also showed a varying RM with reversal, which also indicates a binary origin \citep{Kumar2023}. However, a binary origin is not necessary to interpret the RM fluctuations. As shown by \cite{2022Beniamini}, if the FRB passes through a region with turbulent and even weakly magnetized plasma can result in stochastic RM fluctuations.

Extremely large RMs also have been observed in the vicinity of massive black holes. For example, %a radio emission associated with the Galactic Centre black hole Sagittarius $A^{\ast}$ has RM = $-5\times10^{5}~\rm{rad}~m^{-2}$ , 
PSR J1745-2900, which resides 0.12 pc from Sgr ${A^{\ast}}$ was estimated to have an RM = $-6.69\times10^{4}~\rm{rad}~m^{-2}$ \citep{59}. The RM variation is about 3500 $\rm {rad}~m^{-2}$ \citep{Desvignes2018}. PSR J1746-2850, which is close to Sgr A${^{\ast}}$, is reported to have RM $=-12234~\rm {rad}~m^{-2}$ with variation of $300-400$ $\rm {rad}~m^{-2}$ \citep{2023Abbate}. %\cite{95} also reported a high RM value ($\sim 7000~\rm{rad}~m^{2}$) in the jet components of an active galactic nucleus (AGN) source. 
So it is possible that some FRB sources may be near SMBHs. Combining with the evidence that one magnetar can result in producing of FRB-like bursts, we consider an FRB-generating magnetar orbiting an SMBH, called EMRI. % and FRB can be produced by this magnetar, this is also the Extreme Mass Ratio Inspiral.

%The average value of RM of FRB 20201124A was obtained as $-614~\rm{rad~m}^{-2}$ with a deviation of $\sim16~\rm{rad~m}^{-2}$ and a period of time with constant RM was observed, otherwise 75\% circular polarization was found in this FRB \citep{35,85}, it is proposed that FRB 20201124A may reside in a binary system contains a magnetar \citep{Wang2022}. It has been found that a several hundred $\rm{rad ~m}^{-2}$ variations in RM from FRBs 20181119A, 20190303A, and 20190417A in timescales month, and minor RM variability in FRBs 20181030A, 20190208A, 20190213B and 20190117A in the same timescales. \citep{94}.

EMRI is one of the main sources expected for the future space-borne GW detector LISA to detect \citep{20,21}. When a neutron star (NS) inspirals into an SMBH, the system emits significant amounts of continuous gravitational waves (GWs) along each orbit in the frequency domain to which LISA will be most sensitive. Such GWs can be regarded as `standard siren' if the information about redshift is available. In this case, it is natural to seek the EM counterpart of GW carrying the redshift information. Although it is currently believed that there are no such EM counterparts of EMRI signals in most cases, some models have proposed that EMRI can be associated with EM signals. For example, a new formation channel of EMRIs with tidal disruption flares as EM counterparts was proposed by \cite{57}. Quasi-periodic eruptions generated from the Roche lobe overflows of an evolved star of white dwarf orbiting an SMBH can be detected as EMRI sources with EM counterparts \citep{King2020,58}. Recently, some indications for the EM counterpart of GWs have been found. For instance, GW 190425, a binary NS merger event, was reported to be associated with a one-off FRB 20190425A with the chance probability $2.8~ \sigma$ \citep{60}. However, this association was questioned by \cite{Bhardwaj2023}. \cite{Sridhar2021} and \cite{Katz2023} proposed that periodic FRBs can be produced from the extreme mass ratio binaries that contain ordinary stars and intermediate mass black holes. In this letter, we propose that FRBs may be the EM counterparts of EMRIs.

%Once EMRI signals have been detected, they may provide evidence of a binary system containing a compact star interacting with an SMBH, after being put into operation, LISA and Tianqin are the main detectors for such signals \citep{22,23}. In this letter, we consider this kind of two-body interacting binary system which is capable of producing the low-frequency EMRI signal and FRBs. %The RM and DM are deemed to be contributed by the outflow from the central SMBH. 

%In this work, we consider the FRB-producing magnetar orbiting an SMBH, which is a typical EMRI system producing GWs. Two ways of producing FRB are considered. One is FRB being produced through the cosmic comb process \citep{25}, the other is FRB being produced in the magnetosphere of magnetars. Then we calculated the S/N of gravitational waves for LISA and Tianqin. The event rate of capturing of NS by SMBH is also estimated.

This letter is organized as follows. The relative FRB progenitor models are discussed in Section \ref{sec:FRB progenitor models and observations}. The EMRI signals produced in the NS-SMBH system are given in Section \ref{sec:Model:NS-SMBH binary system}. The event rate is estimated in Section \ref{sec:Event Rate}. Discussions and conclusions are given in Section \ref{sec:Discussion and summary}

\section{FRB progenitor models} \label{sec:FRB progenitor models and observations}

%Although localized FRBs seem to support the magnetar progenitor models \citep{11}, it is too soon to say that all the FRBs are magnetar-origin \citep{12}. 
Since the discovery of FRB 20200428, progenitors related to magnetars have received increasing attention. Several magnetar progenitor models were proposed, for more details see reviews by \cite{13} and \cite{96}. Here we focus only on the single magnetar model proposed by \cite{17} and the cosmic comb model proposed by \cite{25}.

\subsection{Single magnetar model} \label{subsec:Magnetar source models}

The coherent curvature radiation in the magnetosphere may be a possible radiation mechanism of FRBs. In this model, the disturbance near the surface of the magnetar launches an Alfv$\acute{e}$n wave packet, and a fraction of the wave energy is converted into coherent radio radiation. Alfv$\acute{e}$n waves with the non-zero component of transverse wave-vector require an electric current along the magnetic field, which can be supplied by the counter-streaming electron-positron pairs moving at near the speed of light at larger radii as the plasma density decreases with distance from the magnetar surface. The bunch formation is attributed to the two-stream instability. When the Alfv$\acute{e}$n waves reach the radius where electron-positron pairs are insufficient to supply the required current, a strong electric field forms accelerating the particle bunches along the field. The coherent FRB radiation is then produced. This model has been applied to FRB 20200428, which is associated with a magnetar \citep{97,99}

\subsection{Cosmic comb model} \label{subsec:Cosmic comb model}

The cosmic comb is a model taking two-body interaction into consideration to produce FRBs \citep{25}.
In this model, an NS may produce a bright FRB if its magnetosphere is combed by a close, strong plasma stream toward the anti-stream direction. The radiation is detectable only in the inferior conjunction configuration. The plasma stream could be an outflow from an SMBH. In this model, the properties of FRBs are determined by the stream energy flux received by the NS \citep{25}.
Due to the need for the NS magnetic field to be modified, the ram pressure of the stream is required to be greater than the magnetic pressure of the NS magnetosphere. For a non-relativistic and cold flow, the condition can be expressed as

\begin{equation}\label{equ:1}
P_{\mathrm{ram}}>\frac{B^2}{8 \pi}.
\end{equation}

If a dipolar magnetic configuration of the NS is assumed, the magnetic pressure is

\begin{equation}\label{equ:2}
\frac{B_{\mathrm{LC}}^2}{8 \pi}=\frac{B_s^2}{8 \pi}\left(\frac{\Omega R_{s}}{c}\right)^6 \simeq3.4~ \mathrm{erg} ~\mathrm{cm}^{-3} B_{s, 12}^2 P^{-6},
\end{equation}
where $B_{\rm{LC}}$ is the magnetic field strength at the light cylinder,
$B_{\rm{s}}$ and $R_{\rm{s}}$ are the surface magnetic field strength and radius of NS, and $\Omega$ and $P$ are the angular frequency and rotation period of the NS, respectively.
If the above condition is satisfied, the plasma stream arrives and significantly modifies the surface magnetic field of the NS toward the anti-stream direction. After that, magnetic reconnections may be triggered, releasing considerable energy that is adequate to accelerate the particles moving along the magnetic field at relativistic speeds within a short timescale. Eventually, an FRB can be produced through the coherent curvature radiation mechanism. %This model has been applied to FRB 121102 to explain the unique observational properties tentatively \citep{26}.

%\subsection{RM of FRBs} \label{subsec:RM observations of FRB}

%The observed RM is defined by the observed polarization angle of linearly polarized waves as
%\begin{equation}
%\Psi_{\mathrm{obs}}(\lambda)=\Psi_0+\lambda^2 \mathrm{RM},
%\end{equation}
%where $\Psi_{0}$ is the initial polarization angle (PA) of the electromagnetic wave of wavelength $\lambda$. The above formula requires the wave to contain a linear polarization part and the PA should satisfy $\Psi \propto \lambda^{2}$. %In most cases, this condition can be equivalent to \citep{28}
%\begin{equation}
%B \ll B_c=\frac{2 \pi m_e c v}{e} \simeq 360~ \mathrm{G}\left(\frac{v}{1 \mathrm{GHz}}\right),
%\end{equation}
%where $m_{e}$ stands for mass of electron. 
%RM is defined as
%\begin{equation}
%\mathrm{RM}=\frac{e^3}{2 \pi m_e^2 c^4} \int_0^L n_e(l) B_{\|}(l) \mathrm{d} l,
%\end{equation}
%where $n_{e}$ is the number density of electron, and $B_{\|}(l)$ is the parallel component of the magnetic field along the line of sight. 

\section{EMRI signals} \label{sec:Model:NS-SMBH binary system}
We consider a system in which an NS moves along a highly eccentric orbit around the central SMBH, and the orbital pericenter decays owing to the GW radiation, producing the EMRI signals that may be detected by LISA and Tianqin. In such a system, except for the EMRI signals, there are two different ways to produce an FRB. Firstly, the outflow from the SMBH may comb the magnetosphere of NS, producing a bright FRB according to the cosmic comb model. Secondly, an FRB can be produced in the magnetosphere of NS according to the model of \cite{17}.

\subsection{Constraints on orbit}\label{subsec:Constrain on orbit}
It is worth noting that only the scenario an FRB is generated through the cosmic comb model sets limits on the orbit of the binary system. The scenario in which an FRB is produced directly from the magnetosphere without interacting with the SMBH outflow places no restrictions on the orbit. The following discussion in this subsection is only for the cosmic comb scenario.

The mass loss rate of an SMBH can be expressed in the form of Eddington accretion rate $\dot{M}_{\mathrm{Edd}}$ with a dimensionless parameter $f$ reads \citep{28} %which can be used to calculate the ram pressure come from the SMBH \citep{28}.
\begin{equation}\label{equ:3}
\dot{M} = f \dot{M}_{\mathrm{Edd}}=\frac{4 \pi G m_{\rm{p}}}{\epsilon_{\mathrm{BH}} \sigma_T c} f M_{\mathrm{BH}}
 \simeq 2.2 \times 10^{-3} ~\mathrm{M}_{\odot} \mathrm{yr}^{-1} f\left(\frac{M_{\mathrm{BH}}}{10^5 \mathrm{M}_{\odot}}\right).
\end{equation}
We consider an outflow from the SMBH with a typical dimensionless velocity $\beta=v/c=\beta_{-2}/10^{2}$. The ram pressure of the stream at the semi-major axis of orbit $a$ is given by \citep{25}
\begin{equation}\label{equ:4}
P_{\mathrm{ram}} \simeq1.6 ~\mathrm{erg} ~\mathrm{\textrm {cm}^{-3} }\left(\frac{\dot{M}}{M_{\odot} \mathrm{yr}^{-1}}\right) \beta_{-2}\left(\frac{a}{10^{-2} \mathrm{pc}}\right)^{-2}.
\end{equation}
Substituting \equationautorefname~(\ref{equ:2}) and \equationautorefname~(\ref{equ:4}) into \equationautorefname~(\ref{equ:1}), we find that the ram pressure can be modified as
\begin{equation}\label{equ:8}
B_{\rm{s, 13}}^2 P^{-6} \leq 0.46\left(\frac{\dot{M}}{M_{\odot} \mathrm{yr}^{-1}}\right) \beta_{-2}\left(\frac{a}{10^{-2} \mathrm{pc}}\right)^{-2}.
\end{equation}
When the typical parameters of NS (i.e. $B_{s},P$) are given, the comb condition $P_{\rm{ram}}>P_{\rm{B,LC}}$ can be used to constrain the semi-axis $a$
\begin{equation}\label{equ:9}
\frac{a}{10^{-2} \rm{pc}}\leq \left({1.012\times10^{-3} \left(\frac{M_{\mathrm{BH}}}{10^5 ~M_{\odot}}\right) f\beta_{-2} B_{\text {s,13}}^{-2} P^6}\right)^{\frac{1}{2}}.
\end{equation}
For the RM contribution, it can be estimated as \citep{28}
\begin{equation}\label{equ:11}
\begin{aligned}
\mathrm{RM} \sim & \frac{e^3}{2 \pi m_e^2 c^4} B_{\rm{r}} n_{\rm{e}} a \simeq 1.6 \times 10^5~ \mathrm{rad} ~\mathrm{m}^{-2} f  \times\left(\frac{M_{\mathrm{BH}}}{10^5 ~M_{\odot}}\right)\left(\frac{\beta_{-2}}{1}\right)^{-1}\left(\frac{B_r}{1 ~\mathrm{mG}}\right)\left(\frac{a}{10^{-2} ~\mathrm{pc}}\right)^{-1},
\end{aligned}
\end{equation}
where $B_{r}$ is the magnetic field of SMBH at $r\sim a$. We take \equationautorefname~(\ref{equ:9}) as the upper limit of the separation between the NS and the SMBH, and fix $\beta_{-2}$ to $1$. We consider the distances of NSs to be far from the gravitational radius of SMBHs. Therefore, relativistic corrections do not need to be considered here. Combing with \equationautorefname~(\ref{equ:11}), we can estimate the lower limit of RM contributed by the SMBH outflow as 
\begin{equation}\label{equ:10}
    {\mathrm{RM_{lowlim}} = 5.03 \times 10^3 ~\mathrm{rad~m^{-2}}   B_{\mathrm{s, 13}} P^{-3}\left(\frac{f}{10^{-6}}\right)^{\frac{1}{2}}\left(\frac{M_{\mathrm{BH}}}{10^5 M_\odot}\right)^{\frac{1}{2}}\left(\frac{B_{\mathrm{r}}}{\operatorname{1mG}}\right)}.
\end{equation}
The derived RM and DM should be consistent with the observations of those FRBs that may be in the vicinities of SMBHs. This kind of FRB has relatively large RMs and DMs. According to current observations, only FRB 20121102 \citep{75,90} and FRB 20190520B \citep{Dai2022,Anna-Thomas2023} could satisfy this criterion. %FRB 20190520B has not yet been ruled out as being related to massive black holes \citep{Anna-Thomas2023}. 
So we set the accessible range below
\begin{equation}\label{equ:12}
1.6 \times 10^4f\left(\frac{M_{\rm{BH}}}{10^{5} M_{\odot}}\right)\left(\frac{a}{10^{-2}\rm{p c}}\right)^{-1} \geq \rm{RM}_{\rm{lowlim}},
\end{equation}
\begin{equation}\label{equ:13}
{\mathrm{DM} \sim2 f\left(\frac{M_{\rm{BH}}}{10^{5} M_{\odot}}\right)\left(\frac{a}{1 p c}\right)^{-1} \sim \left(10 - 10^3\right)}.
\end{equation}
DM is estimated from the \equationautorefname~(49) in \cite{28}. Besides, in a binary system containing an SMBH, the semi-major axis $a$ is supposed to satisfy the condition that the pericenter $p$, which can be expressed as $p = a \left(1 - e\right )$, should be much larger than the gravitational radius $r_{g}$ to prevent the companion star from falling into the horizon of the SMBH
\begin{equation}\label{equ:14}
a(1-e)\gg r_{g} = \frac{2 G M_{\rm{BH}}}{c^2}.
\end{equation}

EMRI begins with the scattering capture of NS by SMBH, this process tends to result in a high initial eccentricity orbit, which is expected to maintain the GW radiation in the LISA band \citep{42}.
So we take a higher eccentricity range, $0.5 \le e \le 1$. Considering reasonable values of $e$, we can roughly derive
\begin{equation}\label{equ:15}
a\gg\frac{4 G M_{\rm{BH}}}{c^2}.
\end{equation}
Once the value of $f$ is determined, the upper limit value of $a$ can be simply derived by using \equationautorefname~(\ref{equ:9}). Assuming the outflow of the low-luminosity SMBHs have similar properties with the Sgr A* \citep{43}, we set a range of dimensionless parameter $f$ characterizing mass loss rate from $10^{-7}$ to $10^{-2}$ \citep{26}. Here we will take a narrower range because a large value of $f$ (i.e. typically greater than $10^{-4}$) corresponds to an overlarge $\rm{RM}_{lowlim}$, which has never appeared in previous observations. The possible range of the dimensionless parameter $f$ can be estimated as
\begin{equation}
    f\sim \left(1 \times 10^{-7} - 1 \times 10^{-4}\right).
\end{equation}
However, $f=10^{-7}$ corresponds to an extremely short merge time (e.g. $\leq1\rm{yr}$). Hence, we set the value of $f$ from $10^{-6}$ to $10^{-4}$ which is a much more reasonable range.

\subsection{S/N of EMRIs}
The strength of EMRI is determined by the strain amplitude  $h_{0}$. The low-frequency signals in the LISA and Tianqin band will last for many years, which means that the $n^{th}$ harmonic of the binary will spend approximately $f_{n}^{2}/\Dot{{f_{n}}}$ cycles near the fundamental frequency $f_{n}$.

To account for the integration of the accumulated signal, the characteristic strain amplitude is used to represent the amplitude detected by the detector over the entire mission time \citep{44}
\begin{equation}
h_{c, n}^2=\left(\frac{f_n^2}{\dot{f}_n}\right) h_n^2.
\end{equation}
This formula is used to derive the GW strain amplitude at each instantaneous moment
\begin{equation}
h_{c, n}^2=\frac{1}{\left(\pi D_L\right)^2}\frac{2 G}{c^3} \frac{\dot{E}_n}{\dot{f}_n},
\end{equation}
where $D_{\rm{L}}$ is the luminosity distance, $\Dot{E_{\rm{n}}}$ is the power of the $n^{th}$ harmonic, $\Dot{f_{n}}=-2\Dot{P}/P^{2}$ is the chirp rate, and $P$ is the period of the orbit. The S/N can be estimated by \citep{46}
\begin{equation}
(\mathrm{S} / \mathrm{N})^2=\int_{f_{GW1}}^{f_{GW2}} \frac{h_{c}^2(f_{GW})}{h_{n}^2(f_{GW})} d(\ln f_{GW}),
\end{equation}
where $h_{n}^{2}=f_{GW}S_{n}(f_{GW})$ is the spectral density of the detector. The S/N also can be expressed as
\begin{equation}
\mathrm{(S/N)}^2=\frac{h_{c}^2 \Delta f_{GW}}{f_{GW}^2 S_{n}(f_{GW})}.
\end{equation}
The effective strain is defined as \citep{47}
\begin{equation}
h_{\rm{eff}}=h_c \sqrt{\Delta f_{GW} / f_{GW}}.
\end{equation}
The $h_{\rm{eff}}$ can be expressed by $h_{n}$ \citep{51}
\begin{equation}
h_{\rm{eff}}=h_{n} \sqrt{t_{\mathrm{\rm{LISA}}} f_{GW}}=h_n \sqrt{N_{\mathrm{\rm{obs}}}},
\end{equation}
where $N_{\rm{obs}}$ is the number of GW cycles during the whole observing time of LISA, the more cycles that LISA observes the higher the signal-noise ratio will be. We assume the 4 yr mission time of LISA \citep{21}. 
For incoherent search and coherent search of LISA, the threshold of S/N is 20 and $12 - 14$ respectively. Here we set the threshold of the coherent search S/N to be 12. For Tianqin, the threshold of $\rm{S/N}$ is 10 \citep{60}.

\subsubsection{Scenario 1: FRBs produced by cosmic comb model}
For this scenario, FRB is produced by the interaction between the NS and the outflow from SMBH. Therefore, the NS orbit should satisfy the cosmic comb conditions discussed in Section \ref{subsec:Cosmic comb model} and Section \ref{subsec:Constrain on orbit}. Considering a harsh condition to produce FRBs, we assume the companion star is a magnetar. The surface magnetic field strength of magnetar $B_{\rm{s}}$ is fixed to $10^{14}~\rm{G}$ and the period $P$ is fixed to $2~\rm{s}$, which are in the ranges derived from observations \citep{Yuki2010,Esposito2009a,Kargaltsev2012,Kaspi2017}. According to observations, we have little information about $B_{\rm{r}}$ near SMBHs. It has been estimated to be about $1~\rm{mG}$ at $0.12~\rm{pc}$ from Sgr A* \citep{59}. So we choose $B_{\rm{r}}\sim1~\rm{mG}$ for the $10^{6}M_{\odot}$ SMBH. 
We assume the spatial scale $l \propto r_g$, which is related to the inner disk radius with the innermost stable circular orbit. The density of matters in the outflow from SMBH can be estimated as \citep{Porth2010}
\begin{equation}
\rho = \frac{m}{V} \sim \frac{\dot{M} t}{{l}^3} = \frac{f\dot{M}_{\mathrm{Edd}}\frac{l}{v}}{l^{3}} = \alpha \frac{f}{M_{\rm{BH}}},
\end{equation}
where $V$ is the typical spatial volume, $t$ is the typical time scale, and $\alpha$ is the normalized constant. The magnetic field in the outflow is
\begin{equation}\label{equ:21}
B_{\rm{r}} = v_{\rm{A}} (4\pi\rho)^{1/2},
\end{equation}
where $v_{\rm{A}}$ is the Alfv$\acute{e}$n velocity. Eventually, we have the relationship between magnetic field strength and mass of SMBH, i.e., $B_{\rm{r}} \propto f^{0.5} M_{\rm{BH}}^{-0.5}$. We adpot this scaling relationship in our calculations.
The results are shown in Table \ref{tab:Scenario 1} for different cases with different selected parameter values. As discussed before, we fix the mass range of SMBH to $(10^{5}M_{\odot}-10^{7}M_{\odot})$. The range of dimensionless parameter of SMBH mass loss rate $f$ is fixed to $(10^{-6}-10^{-4})$.
We can infer that the maximal horizon of this event is $z\sim0.04$ for the optimistic case. The horizon of detection can be larger if a smaller value of $a$ is reached, but such a system can not exist for enough time to be detected.

\begin{table*}
\centering
\caption{Scenario 1: FRBs produced by the cosmic comb model}
\label{tab:Scenario 1}
\begin{threeparttable}
\begin{tabular}{ccccccccccccc} 
\hline
\hline
Case & $M_{\rm{BH}}$ & $f$ & $B_{\rm{S}}$ & $P_{\rm{S}}$ & $B_{\rm{r}}$ & $a_{\rm{uplim}}$ & $e_{0}$ & $\rm{RM}_{\rm{lowlim}}$ & DM & $z_{\rm{max }}$ & $t_{\rm{merge}}$ \\
& ($10^5 M_{\odot}$) & ($10^{-6}$) & ($10^{14}\rm{G}$) & (s) & (mG) & ($10^{-7}$pc) &  & ($\rm{rad~ m^{-2}}$) & ($\rm{pc~cm^{-3}}$) &   & (yr) \\
\hline
1 & $1$ & $10$ & $1$ & $2$ & $10$ & $2.54$ & $0.5$ & $1.99 \times 10^{5}$ & $25$ & $0.03$ & $ 61$ \\
2 & $10$ & $1$ & $1$ & $2$ & $1$ & $8.05$ & $0.65$ & $1.99 \times 10^{4}$ & $25$ & $0.04$ & $62$  \\
3 & $10$ & $10$ & $1$ & $2$ & $3.2$ & $25.4$ & $0.9$ & $2.01 \times 10^{5}$ & $79$ & $0.01$ & $ 6190$ \\
4 & $10$ & $100$ & $1$ & $2$ & $10$ & $80.5$ & $0.99$ & $1.99 \times 10^{6}$ & $250$ & $0.0005$ & $ 1.56\times10^{7}$ \\
5 & $100$ & $10$ & $1$ & $2$ & $1$ & $80.5$ & $0.99$ & $1.99 \times 10^{5}$ & $250$ & $0.004$ & $ 5890$ \\
\hline
\end{tabular}
\begin{tablenotes}
    % \footnotesize
\item (1) We take the value of spin period $P_{\rm{S}}$ and surface magnetic field strength $B_{\rm{S}}$ of magnetars to 2 s and $10^{14}\rm{G}$ respectively, which are well consistent with observations \citep{Esposito2009a,Yuki2010,Kargaltsev2012,Kaspi2017}. 
(2) $B_{\rm{r}}$ is calculated by combining the \equationautorefname~(\ref{equ:21}) and the given value for the case 2.
(3) DM is estimated by using the \equationautorefname~(\ref{equ:13}). The upper limit of semi-axis $a$ is calculated using \equationautorefname~(\ref{equ:9}).Those limits in \equationautorefname~(\ref{equ:14}) and \equationautorefname~(\ref{equ:15}) should be satisfied.
(4) The lower limit of RM is calculated using \equationautorefname~(\ref{equ:10}).
(5) $t_{\rm{merge}}$ is calculated using equations from  \cite{48}. We fix the initial value of e to $e_{0}=0.5$ and $e_{0}=0.65$ for the top two rows of the table to avoid a too-short merge time.
(6) $z$ is defined as the distance of binary from us when the $\rm{S/N}$ of GW reaches the threshold. We take $a_{\rm{uplim}}$ as the semi-axis, the corresponding $z$ is $z_{\rm{max}}$
\end{tablenotes}
\end{threeparttable}
\end{table*}

\begin{figure*}
\gridline{\fig{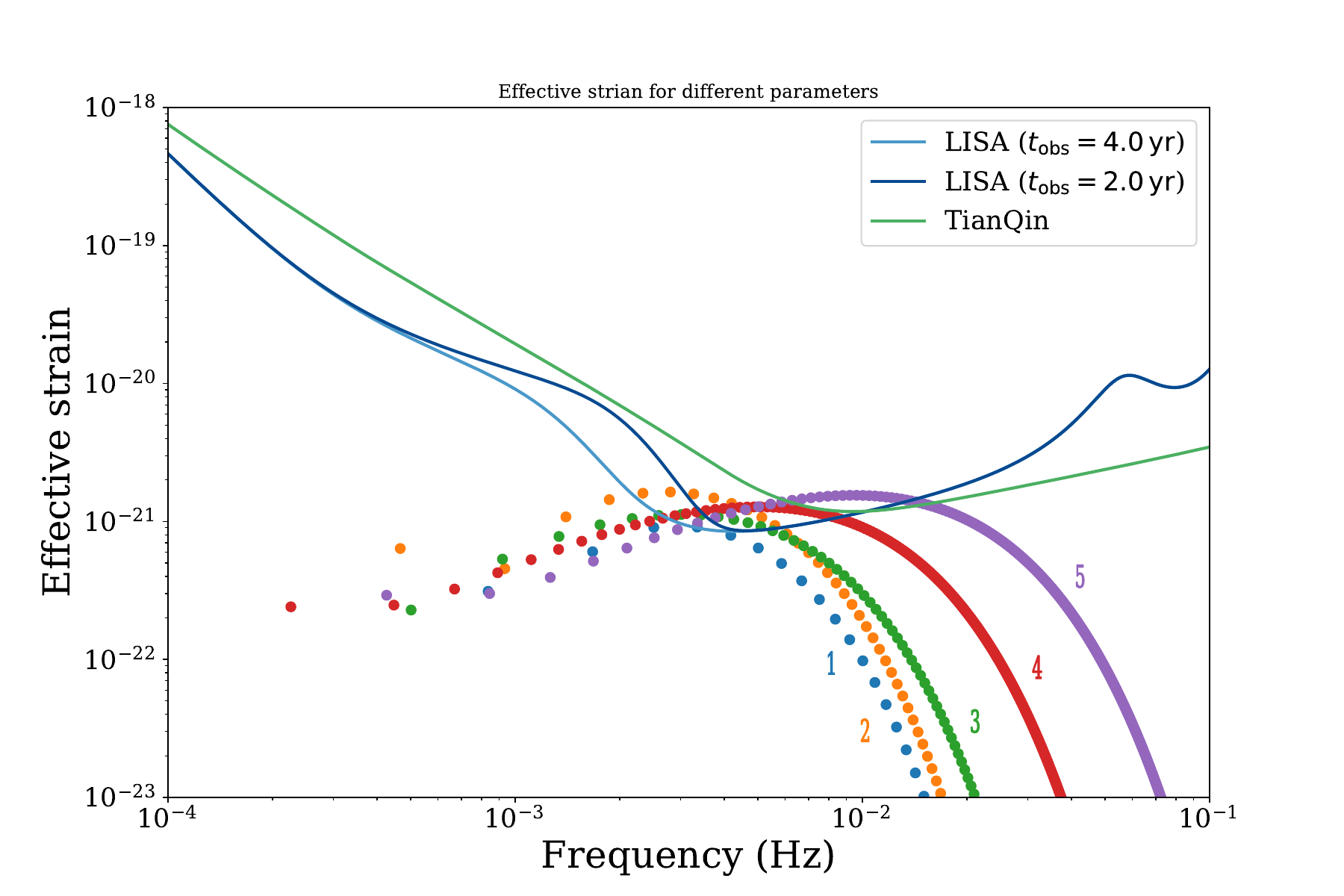}{1\textwidth}{}
          }

\caption{The effective GW strain of EMRI events (dots) and the sensitivity curves $\sqrt{f S_n(f)}$ of LISA and Tianqin (solid lines). Two mission times for LISA are considered \citep{70,69}. The numbers and colors denote different cases listed in Table \ref{tab:Scenario 1}.
\label{fig:Scenario 1}}
\end{figure*}

\begin{figure*}
\gridline{\fig{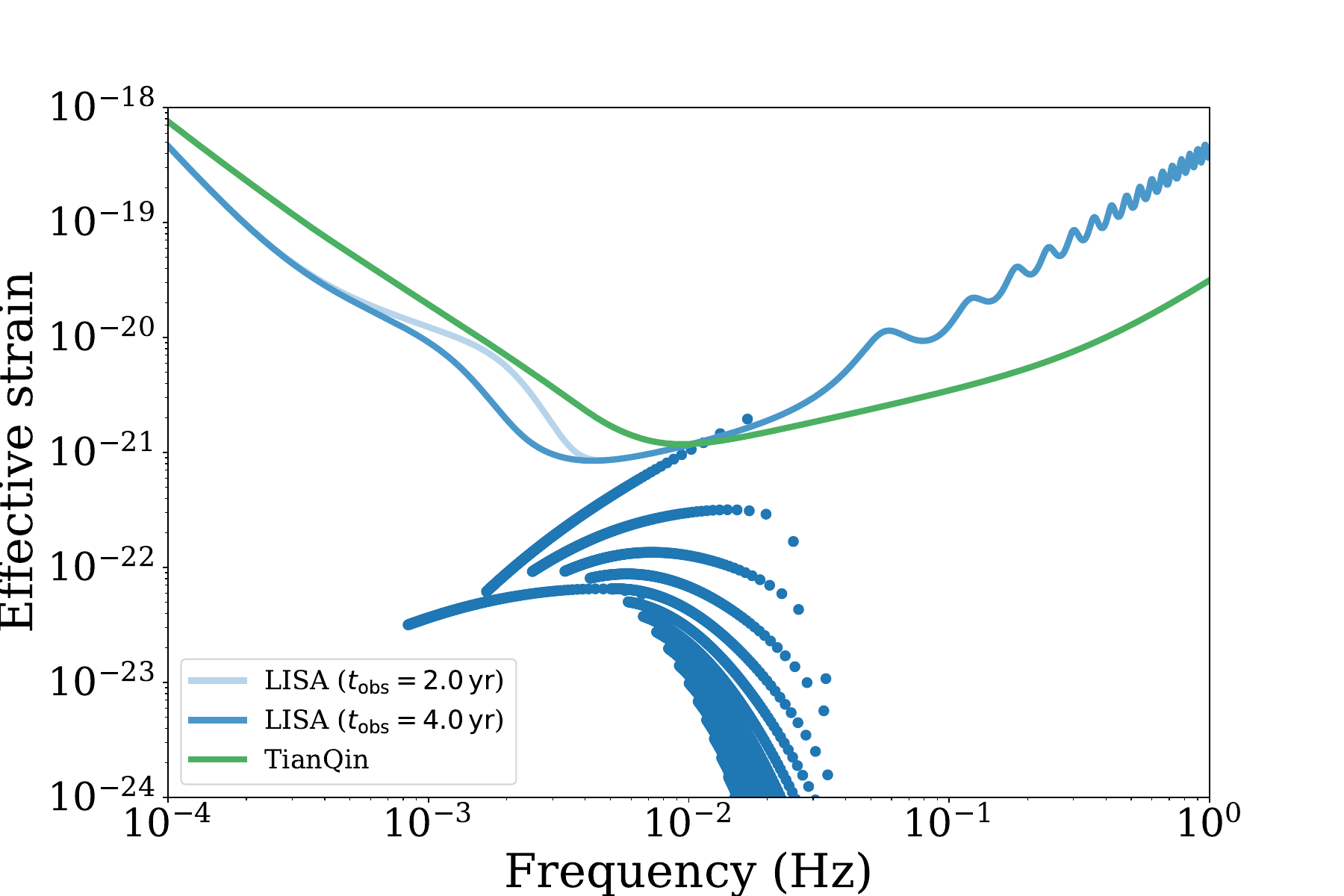}{0.5\textwidth}{(a)}
          \fig{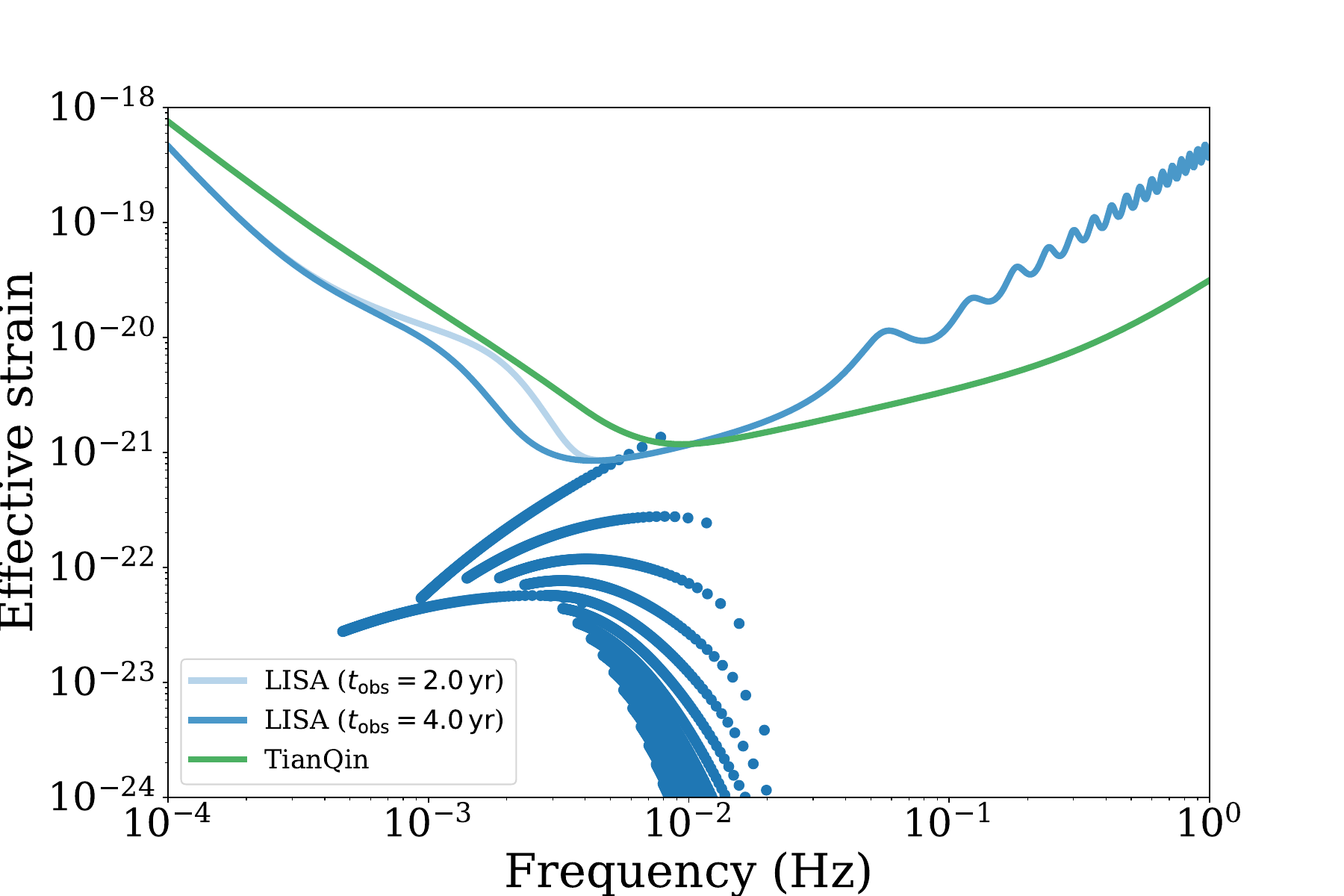}{0.5\textwidth}{(b)}
          }
\caption{The effective GW strain of EMRI events (dots) and the sensitivity curves $\sqrt{f S_n(f)}$ of LISA and Tianqin (solid lines). Two mission times for LISA are considered \citep{70,69}. Panels (a) and (b) correspond to cases 1 and 2 in Table \ref{tab:Scenario 1} when the orbital evolution is considered. The redshift of panel (a) is set to $0.08$, and the redshift of panel (b) is set to $0.2$.
\label{fig:Evolving orbit}}
\end{figure*}

In the above calculations, due to the evolution timescales being much shorter than the observation time, the orbits are considered to be static. Nevertheless, the orbits of the top two cases (i.e. cases 1 and 2) in Table \ref{tab:Scenario 1} only last for less than 62 yr. The evolution of orbits may need to be considered. We show the results in Figure \ref{fig:Evolving orbit}. The GWs may be detected at a further distance (i.e. $z\sim0.2$) or at different frequencies because the orbital parameters are changing with time if the orbits are considered to be evolving.

\subsubsection{Scenario 2: FRBs produced in magnetosphere}
In this situation, FRB is produced from the magnetosphere of the magnetar without interacting with the outflow. The properties of magnetar put no constraints on the orbit, so the semi-axis can be extended to a larger distance than that in scenario 1. We take two possible values of $a$ which can not be too large to retain the frequency of EMRI right in the sensitive band of detectors. The production of FRBs may be related to the magnetic active timescale of magnetars \citep{Margalit2019,Wang2020}, which is about 20 yr for high-mass NSs and 700 yr for normal-mass NSs \citep{Beloborodov2016}. However, the age of SGR J1935 is estimated to be 16,000 yr \citep{Zhou2020}, which is much larger than the typical magnetic active timescale. Therefore, magnetars may emit FRBs over a long period of their lifetime. Furthermore, the dominant mechanism of magnetic field decay of magnetars is poorly constrained \citep{Dall'Osso2012}. The magnetar field decay timescales are constrained between $10^{3}\sim10^{4}$ yr. \cite{2019Beniamini} find the typical magnetar magnetic field decay timescale to be $\sim10^{4}$ yr. From this point of view, SGR J1935 is not unique among the magnetar population. Magnetars may be capable of emitting FRBs in timescales of $10^{3}\sim10^{4}$ yr. Because of the lack of information about the probability that a magnetar produces an FRB, here we suppose that all magnetars are equally likely to produce FRBs. However, only FRB 20200428 has been confirmed to be associated with a magnetar which is not unique compared to the general magnetar population \citep{Margalit2020,Lu2020}. In addition to the magnetars that have been discovered, there may be more undiscovered magnetars in the vicinity of Galactic Center. Besides, in this scenario, the production of FRB and EMRI are not directly connected. Therefore, we can hardly confirm the fraction of the observed FRB population could potentially be ascribed to EMRIs. But the information provided by position, DM and RM may help us exclude some FRBs. The results are shown in Table \ref{tab:Scenario 2}. It can be seen that a larger orbit corresponds to a much smaller detection distance. %Hence, due to the smaller orbits correspond to a short-lived system, there is very few possibilities for detectors to cover a further detection distance.

%For the cosmic comb model, recent general relativity magneto hydrodynamics simulation results point out that the jet of SMBH may not align with the spin axis all the time \citep{43}. The direction of outflow may switch to other directions, and the jet reaching radius $r_{\rm{jet}}$ is also time-varying. An one-off FRB can be generated only once because the outflow changes to other directions or the reaching radius becomes shorter after reaching the magnetar once, no more FRBs are produced later. Although it is widely known that SMBHs reside at the center of massive galaxies such as brightest cluster galaxies (BCGs), the results of simulations and orbital integrations show that SMBH may have an off-centered distance from a few parsecs to hundreds of kiloparsecs \citep{53} which means that it is not necessary for FRB produced in this kind of system from exact the center of galaxies.

\begin{figure*}
\gridline{\fig{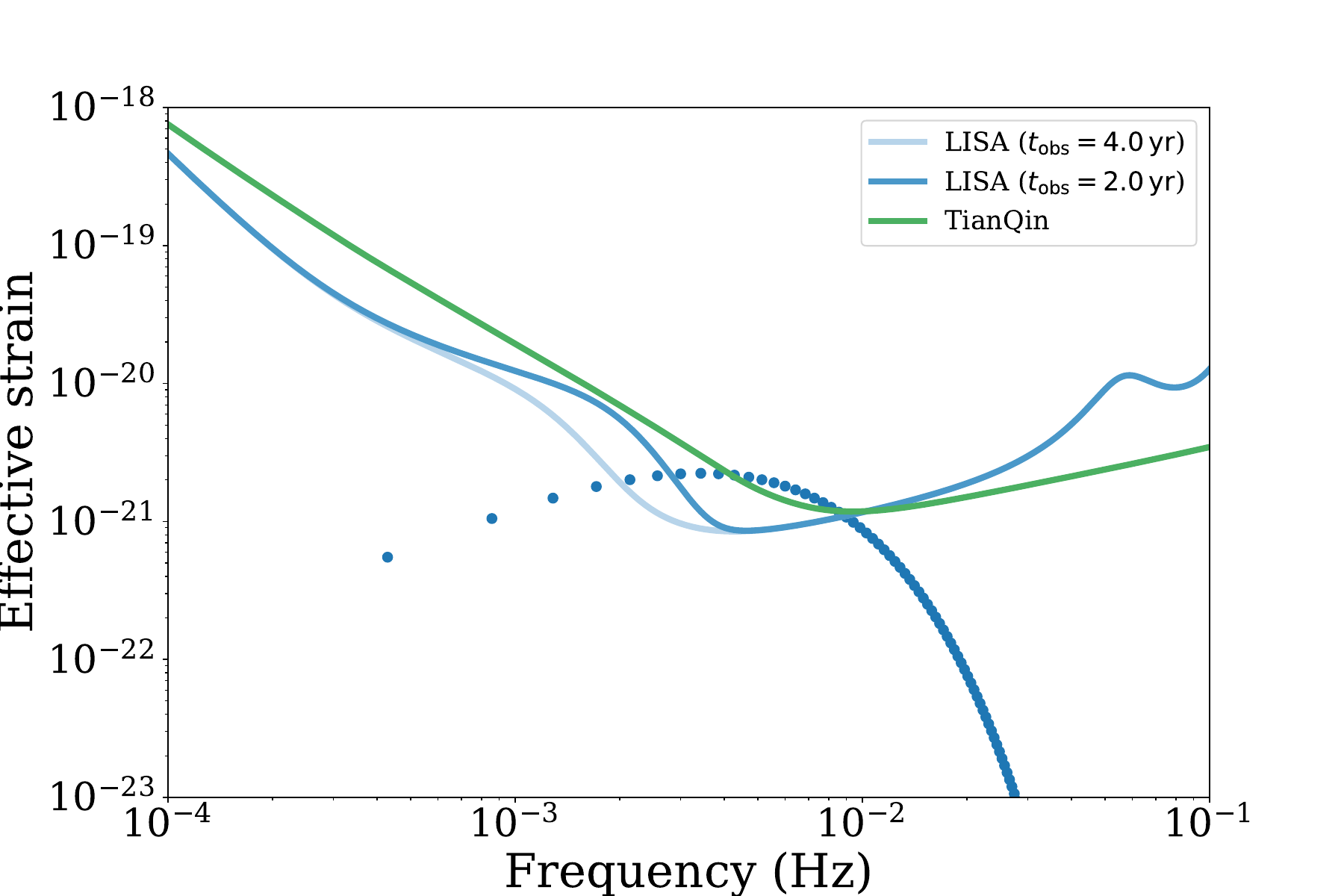}{0.5\textwidth}{(a)}
          \fig{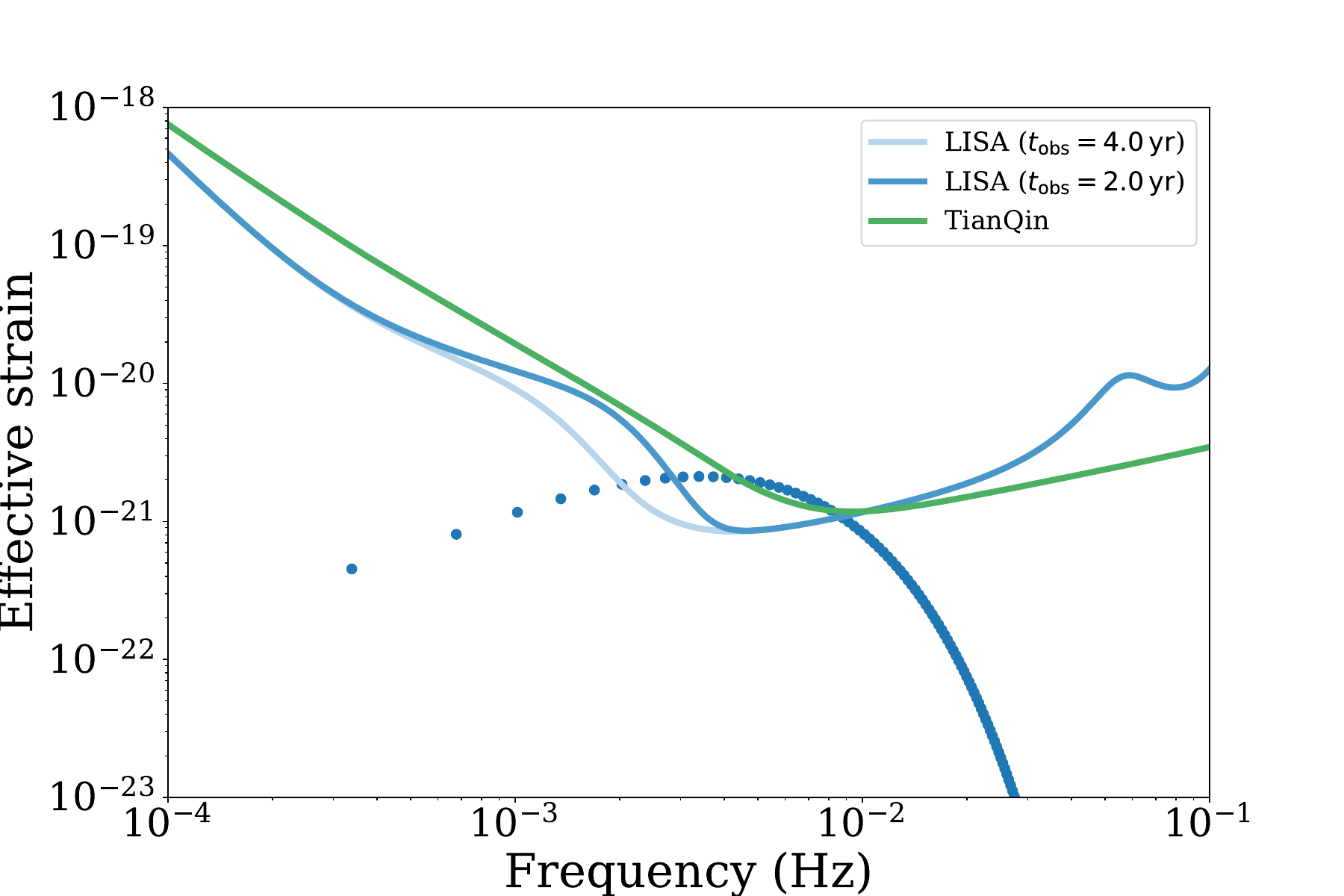}{0.5\textwidth}{(b)}
          }
\caption{The effective GW strain of EMRI events (dots) and the sensitivity curves $\sqrt{f S_n(f)}$ of LISA and Tianqin (solid lines). Two mission times for LISA are considered \citep{70,69}. Panels (a) and (b) correspond to cases 1 and 2 in Table \ref{tab:Scenario 2}, respectively. 
\label{fig:Scenario 2}}
\end{figure*}

\begin{table*}
\centering
\caption{Scenario 2: FRBs produced in magnetosphere}
\label{tab:Scenario 2}
\begin{threeparttable}
\begin{tabular}{cccccccccccc} 
\hline
\hline
Case & $M_{\rm{BH}}$ & $f$ & $B_{\rm{r}}$ & $a$ & $e_{0}$ & $\rm{RM}$ & DM & $z$ & $t_{\rm{merge}}$ & Figure 3\\
 & ($10^5 M_{\odot}$) & ($10^{-4}$) & (mG) & (pc) &  & ($\rm{rad~ m^{-2}}$) & ($\rm{pc~cm^{-3}}$) &   & (yr) & \\
\hline
1 & $100$ & $1$ & $3.2$ & $10^{-4}$ & 0.995 & $5.12 \times 10^{5}$ & $200$ & $0.0003$ & $7200$ & $(a)$ \\
2 & $100$ & $1$ & $3.2$ & $10^{-3}$ & 0.9995 & $5.12 \times 10^{4}$ & $20$ & $0.00001$ & $ 27700$ & $(b)$ \\
\hline
\end{tabular}
\end{threeparttable}
\end{table*}

\section{Event Rate} \label{sec:Event Rate}

In this work, we focus on the binaries composed of an NS and an SMBH. In the single magnetar model, once an FRB-producing magnetar was captured by the SMBH, this magnetar started to enter its inspiral phase. Then it can be counted as an event. In the cosmic comb model, when an NS is captured, it is an event. So we can derive the event rate as follows. The first step is estimating the space density of SMBH in a certain mass range. The second step is estimating the rate at which each SMBH captures NS. The third step is estimating the fraction of magnetars that can produce FRBs among the NS populations. The fourth step is adding together the event rates of the two models to get the total event rate.

The $M-\sigma$ relation is \citep{1}
\begin{equation}
M_{\rm{BH}} = M_{\rm{BH},\ast}\left(\frac{\sigma}{\sigma_{\ast}}\right)^{\lambda},
\end{equation}
where $\lambda=4.72,~M_{\rm{BH},\ast}=3\times10^{6}M_{\odot},~\sigma_{\ast}=90 ~\rm{km~s}^{-1}$, and $\sigma$ represents the velocity dispersion. In addition,  $\lambda=4.02,M_{\rm{BH},\ast}=5\times10^{6}M_{\odot}$ was proposed by \cite{52}. Galaxy luminosity functions can be used to constrain the galaxy velocity dispersion function. Conjoining with the $M_{\rm{BH}}-\sigma$ relation, the black hole mass function is
\begin{equation}
M_{\rm{BH}} \frac{\mathrm{d} N}{\mathrm{~d} M_{\rm{BH}}}=\phi_* \frac{\epsilon}{\Gamma\left(\frac{\gamma}{\epsilon}\right)}\left(\frac{M_{\rm{BH}}}{M_{\rm{BH}}, *}\right)^\gamma \exp \left(-\left(\frac{M_{\rm{BH}}}{M_{\rm{BH}}}\right)^\epsilon\right),
\end{equation}
where $\epsilon=3.08/\lambda$, $\varphi_{\ast}$ is invariable and represents the overall number density of galaxies and $\Gamma(z)$ is the gamma function. The $\varphi_{\ast},M_{\rm{BH},\ast}$and $\gamma$ can be constrained for different types of galaxies \citep{2}. For a typical SMBH mass range, $10^{5}~M_{\odot} \le M_{\rm{BH}} \le 10^{7}~M_{\odot}$, the total spatial density of black hole is approximately as
\begin{equation}
M_{\mathrm{\rm{BH}}} \frac{d N}{d M_{\mathrm{\rm{BH}}}}=2 \times 10^{-3} h_{\rm{70}}^2 ~\mathrm{Mpc}^{-3},
\end{equation}
where $h_{70}\equiv H_{0}/70~\rm{km~ s}^{-1}\rm{Mpc}^{-1}$ is the dimensionless Hubble constant.
The simulation of the Milky Way result gave a prediction that the capture rate of NS is $10^{-6}~\rm{yr}^{-1}$ \citep{3}, and the extreme mass ratio capture rate in a galaxy is 
\citep{4}
\begin{equation}
\frac{1}{2} \frac{M_{\mathrm{df}}(m)}{m t} \approx \frac{F M_{\rm{BH},\ast}}{2 m t}\left(\frac{M_{\rm{BH}}}{M_{\rm{BH},\ast}}\right)^{\frac{3}{8}}\left(\frac{m}{M_{\odot}}\right)^{\frac{1}{2}} \approx 10^{-4} F\left(\frac{M_{\rm{BH}}}{M_{\rm{BH},\ast}}\right)^{\frac{3}{8}}\left(\frac{m}{M_{\odot}}\right)^{-\frac{1}{2}} \mathrm{yr}^{-1},
\end{equation}
where $F$ is the fraction of the total stellar mass in NS, and $t\sim10^{10}~\rm{yr}$. Fixing $m=1.4~M_{\odot}$ and $F=0.01$, we can derive the event rate $R$, which is shown in Table \ref{tab:Calculation result}. The typical value is about 1 $\rm{Gpc^{-3}~yr^{-1}}$. This is the event rate of cosmic comb model. 

Then we present the estimation for the event rate of single magnetar model. We firstly adopt the result that the fraction of NSs being born as magnetars is $\sim10\%$ \citep{2010Popov}, but \cite{2019Beniamini} also estimated the fraction to be $\sim40\%$. Secondly, we consider the requirement that the FRB-producing magnetar should be in its active lifetime when it is captured. The active lifetime that the magnetar spends in the EMRI inspiral phase during its whole life can be roughly calculated. The age of an NS can easily reach Gyr, and the active phase only lasts for $\sim10^{4}~\rm{yr}$. So the fraction of active magnetars can be calculated as $10^{4}\rm{yr}/10^{9}\rm{yr}\sim10^{-5}$. In such a situation, the event rate is so tiny: $10^{-6}~\rm{Gpc^{-3}~yr^{-1}}$ or $4\times10^{-6}~\rm{Gpc^{-3}~yr^{-1}}$. %If the existing observational samples of magnetar can be considered as a good sample of the magnetar populations in the universe, we can use the observations to estimate the fraction of magnetars that satisfy the magnetic field strength requirement. Five of the 24 confirmed magnetars have magnetic field strength lower than $10^{14}~\rm{G}$ \citep{Olausen2014}, so we can simply assume the fraction of magnetars that can not emit FRBs is $5/24\sim 20\%$. In other words, the fraction of magnetars with field strength greater than $10^{14}~\rm{G}$ is estimated to be $\sim80\%$ which constraints the event rate to $\sim0.08~\rm{Gpc^{-3}~yr^{-1}}$ or $\sim0.32~\rm{Gpc^{-3}~yr^{-1}}$. Under such circumstances, the event rate is much higher than the former tiny one. Nevertheless, the selection effect is neglected.} 

We should add together the event rates corresponding to the two models to get the total event rate. Eventually, the total event rate is estimated to be $\sim1~\rm{Gpc^{-3}~yr^{-1}}$, which is similar to the event rate of the cosmic comb model. This is because the contribution from the cases that FRBs produced in magnetosphere is so tiny. Captures of compact objects by SMBHs serve as the standard formation channel of EMRIs. Other processes including the tidal separation of compact binaries and formation or capture of NSs in accretion disks can also form EMRIs \citep{2007Amaro-Seoane,2018Maggiore}. Beyond that, `fake plunges' \citep{2013Amaro-Seoane} and Hills binary disruption \citep{2019Sari} also contribute to the event rate of EMRIs significantly. Therefore, the event rate is not grossly overestimated.

Based on the detection distances of the detectors shown in Table \ref{tab:Scenario 1} and Table \ref{tab:Scenario 2}, few events fall within the detection range in a 4-yr mission time of LISA.
However, there are uncertainties in both the spatial density of SMBHs and the capture rate. The SMBH space density in our range was found to be magnitude lower using SDSS data \citep{5}. However this extrapolation should not be reliable on account of the insufficient resolution of SDSS spectra.

\begin{table*}
\centering
\caption{Estimated rate $R$.}
\label{tab:Calculation result}
\begin{tabular}{cc} 
\hline
\hline
$M_{\rm{BH}}(M_{\odot})$ & R ($\rm{Gpc^{-3}~yr^{-1}}$) \\
\hline
$10^{5}$ & $0.47$  \\
$10^{6}$ & $1.12$  \\
$10^{7}$ & $2.65$  \\
\hline
\end{tabular}
\end{table*}

\section{Discussion and conclusions} \label{sec:Discussion and summary}
The host galaxies of a few FRBs were confirmed by arcsecond-scale localization by modern radio interferometers \citep{9}. Recently, the potential host galaxies of FRB 20200223B and FRB 20190110C were proposed \citep{10}. These observations found that most of FRBs are not located in the centers of their host galaxies. However, those FRBs with less than a few kpc offset can still be potential candidates for such systems because SMBHs may have an offset of kpc scale from the center of galaxies \citep{Reines2020}. 

%In this work, we fix various parameters, for example, dimensionless wind speed $\beta$, period $P_{\rm{s}}$ and surface magnetic field strength $B_{\rm{s}}$ of the companion star and the magnetic field strength in the vicinity of SMBH $B_{\rm{r}}$. There are significant uncertainties in these parameters that may lead to impacts on the results. Nevertheless, we only aim to propose FRB can be the electromagnetic counterpart to EMRI, and through our analysis to confirm this event may be detected by space GW detector LISA and Tianqin. \cite{101} expects that FRB may be produced by G-Z effect, at the same time low-frequency GW may be generated from the same location. But this may differ from what we considered here because of the different origin of the low-frequency GW .

For scenario 1, due to the limits of the cosmic comb model, the magnetar must approach the SMBH closely (e.g. $r\sim10^{-6}~\rm {pc}$). We use the upper limit of $a$ to calculate the detection range. Theoretically, we can detect the event with a smaller semi-axis $a$ as long as the lower limit of the orbit is met. However, binary systems with extremely small semi-axis are actually rare in the universe. For scenario 2, when the semi-axis $a$ is the order of $10^{-3}~\rm{pc}$, it seems hard to detect the EMRI signals. Unless the orbit has a large eccentricity (e.g.$~e>0.995$), the peak of signal amplitude can reach the most sensitive frequency band of the detectors. In general, the maximum horizon of such events is $z\sim0.04$ in the most optimistic case. If the magnetar gets closer to the SMBH, it can be extended to a further detection distance but is not likely to be detected due to the short lifetime. Under the assumption that cosmic comb model proves to be a viable model for FRB production, the total event rate can be as high as 1 Gpc$^{-3}$ yr$^{-1}$.

Even though there are various potential candidates for FRB progenitors, the origin of FRBs still remains mysterious. In the future, the detection of EMRIs with FRBs as the EM counterpart will help us to study FRBs. These sources are also important for cosmological applications, such as measuring the Hubble constant \citep{Abbott2017,YuH2018,Wu2022}. 

\section*{acknowledgements}
We thank the anonymous referee for constructive comments. We thank Yuan Feng, Chen Xian, Chen-Ran Hu, Xuan-Dong Jia, Hao-Tian Lan, Jian-Guo He and Yun-Qing Wang for their helpful discussions. This work was supported by the National Natural Science Foundation of China (grant Nos. 12273009,  12041304 and 12288102), the National SKA Program of China (grant No. 2022SKA0130100), and the China Manned Spaced Project (CMS-CSST-2021-A12).

\bibliography{sample631}{}
\bibliographystyle{aasjournal}

\end{document}